\documentclass[twocolumn,showpacs,preprintnumbers,amsmath,amssymb,superscriptaddress,nofootinbib]{revtex4}
\usepackage{epsfig}
\usepackage{psfrag}
\usepackage{amsfonts}
\usepackage{graphicx}
\usepackage{dcolumn}
\usepackage{bm}

\begin{document}


\title{The Subtleties of the Wigner Function Formulation of the Chiral Magnetic Effect}

\author{Yan Wu}
\affiliation{
School of Mathematice and Physics, China Geoscience University(Wuhan), Wuhan 430074, China
}
\affiliation{
Institute of Particle Physics and Key Laboratory of Quark and Lepton Physics
(MOE),  Central China Normal University, Wuhan 430079, China}

\author{De-fu Hou}
\affiliation{
Institute of Particle Physics and Key Laboratory of Quark and Lepton Physics
(MOE),  Central China Normal University, Wuhan 430079, China}

\author{Hai-cang Ren}
\affiliation{
Physics department, The Rockefeller University, 1230 York Avenue, New York, New York 10021-6399, USA}
\affiliation{
Institute of Particle Physics and Key Laboratory of Quark and Lepton Physics
(MOE),  Central China Normal University, Wuhan 430079, China}

\date{\today}

\begin{abstract}

We assess the applicability of the Wigner function formulation in its present form to the chiral Magnetic Effect and noted some issues regarding
the conservation and the consistency of the electric current in the presence of an inhomogeneous and time dependent axial chemical potential.
The problems are rooted in the ultraviolet divergence of the underlying field theory associated with the axial anomaly and can be fixed
with the Pauli-Villars regularization of the Wigner function.

\end{abstract}

\maketitle



\section{Introduction}

The chiral magnetic effect of electrically charged fermions, proposed in \cite{Kharzeev1,Kharzeev2,Fukushima, Kharzeev3}, remains a subject being actively investigated. Because of the axial anomaly, a nonzero axial charge density
controlled by the chemical potential $\mu_5$ in an constant magnetic field {\bf B} generates an electric current along the magnetic field, which takes the form
\begin{equation}
{\bf J}=\frac{e^2}{2\pi^2}\mu_5{\bf B}
\label{cme}
\end{equation}
for a constant $\mu_5$ and is expected to be free from higher order corrections because of the non-renormalization theorem of the anomaly. A net axial charge density can emerge via topological
charge fluctuations of QCD in a quark-gluon plasma or via topological surface modes of certain Weyl semi-metals. The experimental indications of CME include the charge separation
post off-central heavy ion collisions in RHIC\cite{ChargeSep1} and the negative magnetoresistance of a Weyl semi-metal in a magnetic field\cite{SemiMetal}.

The experimental situation is far from ideal in RHIC, where the magnetic field generated via off-central collision is inhomogeneous and transient,
and the thermal equilibrium, if realized, is local. More theoretical investigations of CME are still required to enrich its phenomenological predictions and solidify its
experimental evidences observed so far. The field theoretic method \cite{Fukushima,Kharzeev4,Hou,Landsteiner1}, holography \cite{Yee,Latrak} and kinetic theories \cite{Gao} are the three main approaches explored in the
literature.

The kinetic theory is particularly suitable to describe a system not in a global thermal equilibrium. At its very center lies the Wigner function that links various hydrodynamic
quantities of the system to the Green's functions of the underlying quantum field theory. The Wigner function was introduced in RHIC physics in \cite{Vasak} and applied
recently to the QGP with net axial charge density \cite{Gao}. Among its successes are the reproduction of the chiral magnetic current, chiral vortical current and axial anomalies
obtained from the field theoretic approaches at a global thermal equilibrium, i.e., at a constant tempearture and axial chemical potential. What is remarkable is the absence of an explicit UV regularization,
which is the underlying mechanism of the axial anomaly.
The more interesting situation away from a global thermal equilibrium, say, with an inhomogeneous and time dependent axial chemical potential is beyond the formulation employed
in \cite{Gao}.

In this work, we would like to explore the Wigner function from field theoretic perspectives. We point out a subtlety of the Wigner function being used for
CME because of the ultraviolet ambiguity hidden in its formulation.
While the subtlety does not impact on the existing result of \cite{Gao}, where the axial chemical potential is assumed constant, a number of
problems, including the violation of the current conservation and consistency, show up when the axial chemical potential becomes inhomogeneous and/or time dependent,
It implies that the Wigner function in its present form is not complete yet to serve its purpose of describing a non-equilibrium thermodynamics and UV regularization is necessary.
Including the Pauli-Villars regularization in the Wigner function formalism, the electric
current extracted is conserved and consistent for an arbitrarily spacetime dependent axial vector potential. In addition, the regularized Wigner function displays
the same sensitivity to the order of the limit when the axial chemical potential $\mu_5$ approaches to a constant noted before \cite{Hou}: The full CME current (\ref{cme}) is recovered
when the time dependence is switched after removing spatilly homogeneity. A different form of the current emerges if the order of the limit is reversed and vanishes at thermal equilibrium.

This paper is organized as follows: The UV ambiguity of the un-regularized Wigner function is discussed in the next section. The chiral magnetic current with a PV regularized
Wigner function is calculated in the section III. The section IV concludes the paper.
Throughout this paper, we shall stay with the Euclidean metric $ds^2=dx_\mu dx_\mu$ with all $\gamma$ matrices hermitian and $x_4=it$ for a real time $t$. The closed-time-path (CTP) Green function are employed for the field theoretic calculations since it applicable in
both equilibrium and non-equilibrium situations.

\section{The UV problems with the unregularized Wigner function}

Following \cite{Gao}, the Wigner function for Dirac fermions at the phase space point $(x,p)$ is a $4\times 4$ matrix with its elements defined by
\begin{equation}
W_{\alpha\beta}(x,p)=\int\frac{d^4y}{(2\pi)^4}e^{-ip\cdot y}<\bar\psi_\beta(x_+)U(x_+,x_-)\psi_\alpha(x_-)>,
\label{wigner}
\end{equation}
where the gauge link
\begin{equation}
U(x_+,x_-)=e^{ie\int_{x_-}^{x_+}d\xi_\mu A_\mu(\xi)}
\end{equation}
with $x_\pm=x\pm\frac{y}{2}$ and $A_\mu$ the gauge potential. The symbol $<...>$ denotes an ensemble average, which is not necessarily in a global thermal equilibrium.
The Heisenberg equations of motion of field operators involved lead to a set of c-number equations satisfied by $W(x,p)$, which for the simple case without interaction, can be
solved explicitly. The electric current density can be extracted formally from the solution $W(x,p)$ according to
\begin{eqnarray}
J_\mu(x)&=&ie\int\frac{d^4p}{(2\pi)^4}{\rm tr}W(x,p)\gamma_\mu\nonumber\\
&=&ie\int d^4y\delta^4(y)U(x_+,x_-)<\bar\psi(x_+)\gamma_\mu\psi(x_-)>\nonumber\\
&=&\lim_{y\to 0}J_\mu(x,y)
\label{limit}
\end{eqnarray}
with
\begin{equation}
J_\mu(x,y)=ieU(x_+,x_-)<\bar\psi(x_+)\gamma_\mu\psi_\beta(x_-)>,
\label{regcurrent}
\end{equation}
but the UV divergence embedded in the operator product at the same point makes the limit ill-defined.

To illustrate the problem, we consider a massless Dirac field in an external electromagnetic field $A_\mu$ and an axial vector field $A_{5\mu}$ with the action
\begin{equation}
S=\int dt\int d^3x{\cal L},
\label{action}
\end{equation}
where the Lagrangian density is given by
\begin{equation}
{\cal L}=-\bar\psi\gamma_\mu(\partial_\mu-ieA_\mu-i\gamma_5A_{5\mu})\psi
\label{lagrangian}
\end{equation}
The axial chemical potential $\mu_5$ in (\ref{cme}) corresponds to the temporal component of the axial vector field, i.e.
$A_{5\mu}=({\bf A}_5,-i\mu_5)$. The spatial components are relevant if the axial magnetic effect is related to the topological fluctuation in QCD via
$A_{5\mu}=\partial_\mu\theta$.

The field theoretic method to calculate the ensemble average in (\ref{wigner}) is the closed time path Green function formation which was proposed
in \cite{Keldysh, Martin} and was systematically developed in \cite{Chou}. The time integral of the action, (\ref{action}) consists of two branches, one from $-\infty$ to $\infty$
and the other from $\infty$ to $-\infty$ and degrees of freedom are thereby doubled. As a result, all field variables acquire an additional index labeling
the time branch where they are defined. Consequently, a fermion propagator becomes
\begin{eqnarray}
{\cal S}_{\rm CTP}(x,y)=\left(\begin{array}{cc}{\cal S}_{11}(x,y) & {\cal S}_{12}(x,y)\\
                                      {\cal S}_{21}(x,y) & {\cal S}_{22}(x,y)\\ \end{array}\right)
                                      \end{eqnarray}
where each block is a 4 by 4 matrix in Dirac space. We have
\begin{eqnarray}
{\cal S}_{11}(x,y)_{\alpha\beta} &\equiv& <T[\psi_\alpha(x)\bar\psi_\beta(y)]>\nonumber\\
{\cal S}_{12}(x,y)_{\alpha\beta} &\equiv& -<\bar\psi_\beta(y)\psi_\alpha(x)>\nonumber\\
{\cal S}_{21}(x,y)_{\alpha\beta} &\equiv& <\psi_\alpha(x)\bar\psi_\beta(y)>\nonumber\\
{\cal S}_{22}(x,y)_{\alpha\beta} &\equiv& <\tilde T[\psi_\alpha(x)\bar\psi_\beta(y)]>\nonumber\\
\label{definition}
\end{eqnarray}
where $x=(\mathbf{x},it)$, $T$ denotes time ordering and $\tilde T$ anti-time ordering. An identity,
\begin{equation}
{\cal S}_{11}(x,y)+{\cal S}_{22}(x,y)-{\cal S}_{12}(x,y)-{\cal S}_{21}(x,y)=0
\label{identity}
\end{equation}
follows from the definition (\ref{definition}). It is straightforward to link the LHS of (\ref{regcurrent})
to different components of the CTP propagator, i.e.
\begin{eqnarray}
J_\mu(x,y) &=& -\frac{ie}{2}{\rm tr}[{\cal S}_{12}(x_-,x_+)+{\cal S}_{21}(x_-,x_+)]\gamma_\mu\nonumber\\
&=& \frac{ie}{2}{\rm tr}[{\cal S}_{11}(x_-,x_+)+{\cal S}_{22}(x_-,x_+)\gamma_\mu\nonumber\\&=&\frac{ie}{2}{\rm Tr}{\cal S}(x_-,x_+)\gamma_\mu
\end{eqnarray}
with the trace ${\rm tr}$ acting on Dirac indices and ${\rm Tr}$ including the CTP indices as well.
The expansion of the propagator ${\cal S}_{ab}(x_-,x_+)$ to the linear power of the gauge potential $A_\mu$ and the axial vector
field $A_{5\mu}$ reads
\begin{eqnarray}
&&{\cal S}_{ab}(x_-,x_+) = S_{ab}(x_-,x_+)\nonumber\\&-&\sum_c\int d^4zS_{ac}(x_--z)\gamma_{\rho5}^c S_{cb}(z-x_+)A_{5\rho}(z)\nonumber\\
&-& e\sum_c\int d^4zS_{ac}(x_--z)\gamma_{\rho}^c S_{cb}(z-x_+)A_{\rho}(z)\nonumber\\
&+& e\sum_{cd}\int d^4z_1\int d^4z_2S_{ad}(x_--z_2)\gamma_{\lambda5}^d\nonumber\\ &&\times S_{dc}(z_2-z_1)\gamma_\rho^cS_{ca}(z_1-x_+) A_\rho (z_1) A_{5\lambda}(z_2) \nonumber\\
&+& e\sum_{cd}\int d^4z_1\int d^4z_2S_{ac}(x_--z_2)\gamma_{\rho}^c\nonumber\\ &&\times S_{cd}(z_2-z_1)\gamma_{\lambda5}^dS_{da}(z_1-x_+)A_\rho (z_2) A_{5\lambda}(z_1)
\label{expansion}
\end{eqnarray}
with $\gamma_{\mu}^1=\gamma_{\mu}$, $\gamma_{\mu}^2=-\gamma_{\mu}$, $\gamma_{\mu5}^1=\gamma_{\mu}\gamma_5$ and $\gamma_{\mu5}^2=-\gamma_{\mu}\gamma_5$, where $S_{ab}(x-y)$ is the free Dirac propagator. Substituting (\ref{expansion}) and the expansion
\begin{equation}
U(x_+,x_-)=1+ie\int_{x_-}^{x_+}d\xi_\nu A_\nu(\xi)+O(A^2)
\end{equation}
into (\ref{regcurrent}) and making appropriate Fourier transformations, we obtain that
\begin{eqnarray}
J_\mu^a(x,y)&=&e^2\int\frac{d^4q_1}{(2\pi)^4}\int\frac{d^4q_2}{(2\pi)^4}e^{i(q_1+q_2)\cdot x}\nonumber\\ &&\times\Lambda_{\mu\rho\lambda}^{abc}(q_1,q_2)
A_\rho^b(q_1)A_{5\lambda}^c(q_2)
\label{JLambda1}
\end{eqnarray}
with the kernel
\begin{eqnarray}
\Lambda_{\mu\rho\lambda}^{abc}(q_1,q_2)&=&-y_\rho\delta_{ab}K_{\mu\lambda}^{ac}(q_2)\nonumber\\&&
-i[K_{\mu\rho\lambda}^{(1)abc}(q_1,q_2)+K_{\mu\rho\lambda}^{(2)abc}(q_1,q_2)]\nonumber\\
\label{kernel1}
\end{eqnarray}
where
\begin{eqnarray}
K_{\mu\lambda}^{ab}(q) &=& e^{-\frac{i}{2}q\cdot y}\int\frac{d^4p}{(2\pi)^4}e^{-ip\cdot y}{\rm tr}\gamma_\mu^aS_{ab}(p+q)\gamma_{\lambda5}^bS_{ba}(p)\nonumber\\
K_{\mu\rho\lambda}^{(1)abc}(q_1,q_2) &=& e^{-\frac{i}{2}(q_1+q_2)\cdot y}\int\frac{d^4p}{(2\pi)^4}e^{-ip\cdot y}\nonumber\\ &&
\times{\rm tr}\gamma_\mu^aS_{ac}(p+q_1+q_2)\gamma_{\lambda5}^cS_{cb}(p+q_1)\gamma_\rho^bS_{ba}(p)\nonumber\\
K_{\mu\rho\lambda}^{(2)abc}(q_1,q_2) &=& e^{-\frac{i}{2}(q_1+q_2)\cdot y}\int\frac{d^4p}{(2\pi)^4}e^{-ip\cdot y}\nonumber\\ &&
\times{\rm tr}\gamma_\mu^aS_{ab}(p+q_1+q_2)\gamma_{\rho}^bS_{bc}(p+q_2)\gamma_{\lambda5}^cS_{ca}(p)\nonumber\\
\label{kernel2}
\end{eqnarray}
and the repeated CTP indices in (\ref{kernel1}) and (\ref{kernel2}) are not to be summed.
The momentum representation of various components of the free CTP fermion propagator are given explicitly by
\begin{eqnarray}
S_{11}(p|m) &=& \frac{i}{p\!\!\!/+i0^+-m}-\pi\frac{p\!\!\!/+m}{E}\nonumber\\ &&
\times[f(E)\delta(p_0-E)+f(E+\mu)\delta(p_0+E)]\nonumber\\
S_{12}(p|m) &=& -\pi\frac{p\!\!\!/+m}{E}\lbrace f(E)\delta(p_0-E)\nonumber\\ &&
+[f(E)-1]\delta(p_0+E)\rbrace\nonumber\\
S_{21}(p|m) &=& -\pi\frac{p\!\!\!/+m}{E}[f(E)-1]\delta(p_0-E)\nonumber\\ &&
+f(E)\delta(p_0+E)\rbrace\nonumber\\
S_{22}(p|m) &=& \frac{-i}{p\!\!\!/-i0^+-m}-\pi\frac{p\!\!\!/+m}{E}
[f(E)\delta(p_0-E)\nonumber\\ &&+f(E)\delta(p_0+E)]
\end{eqnarray}
where $p=(\mathbf{p},ip_0)$, $E=\sqrt{{\mathbf p}+m^2}$, $p\!\!\!/\equiv -i\gamma_\nu p_\nu$ and $f(x)$ is the single particle distribution function.
At thermal equilibrium $f(E)=\frac{1}{e^{\beta E}+1}$ with $\beta$ the inverse temperature. For the purpose of the Pauli-Villars regularization to be discussed
in the next section, we indicate explicitly the dependence on mass $m$. For the massless propagators of this section, $S_{ab}(p)\equiv S_{ab}(p|0)$

Now we show two problems coming from the limit in (\ref{limit}):

$\ \ \ \ \ \ \ \ \ \ \ \ \ \ \ \ \ \ \ \ \ \ \ \ \ \ \ \ \ \ \ \ \ \ \ \ \ \ \ \ \ \ \ \ \ \ \ \ \ \ \ \ \ \ \ \ \ \ \ \ \ \ \ \ \ \ \ \ \ \ \ \ \ \ $
\noindent{\it 1. The nonconservation of the electric current}

Taking the divergence of (\ref{JLambda1}), we have
\begin{eqnarray}
\frac{\partial}{\partial x_\mu}J_\mu^a(x,y) &=& ie^2\int\frac{d^4q_1}{(2\pi)^4}\int\frac{d^4q_2}{(2\pi)^4}e^{i(q_1+q_2)\cdot x}(q_1+q_2)_\mu\nonumber\\ &&\times\Lambda_{\mu\rho\lambda}^{abc}(q_1,q_2)
A_\rho^b(q_1)A_{5\lambda}^c(q_2),
\label{dLambda}
\end{eqnarray}
with
\begin{eqnarray}
&&(q_1+q_2)_\mu\Lambda_{\mu\rho\lambda}^{abc}(q_1,q_2)\nonumber\\
&=& -y_\rho\delta_{ab}(q_1+q_2)_\mu K_{\mu\lambda}^{ac}(q_2)\nonumber\\ &&
-i(q_1+q_2)_\mu[K_{\mu\rho\lambda}^{(1)abc}(q_1,q_2)+K_{\mu\rho\lambda}^{(2)abc}(q_1,q_2)]\nonumber\\
\label{divLambda}
\end{eqnarray}
Using the following identities of the free CTP propagator,
\begin{eqnarray}
S_{11}(p+q)q\!\!\!/S_{11}(p) &=& i[S_{11}(p)-S_{11}(p+q)]\nonumber\\
S_{21}(p+q)q\!\!\!/S_{11}(p) &=& -iS_{21}(p+q)\nonumber\\
S_{11}(p+q)q\!\!\!/S_{12}(p) &=& iS_{12}(p)\nonumber\\
S_{21}(p+q)q\!\!\!/S_{12}(p) &=& 0.
\label{ward}
\end{eqnarray}
and shifting some of the integration momenta, we find
\begin{eqnarray}
&&(q_1+q_2)_\mu[K_{\mu\rho\lambda}^{(1)111}(q_1,q_2)+K_{\mu\rho\lambda}^{(2)111}(q_1,q_2)]\nonumber\\
&=& (e^{-\frac{i}{2}(q_1-q_2)\cdot y}-e^{-\frac{i}{2}(q_1+q_2)\cdot y})\nonumber\\ &&\times\int\frac{d^4p}{(2\pi)^4}e^{-ip\cdot y}
{\rm tr}\gamma_\rho S_{11}(p)\gamma_{\lambda}\gamma_5 S_{11}(p+q_1)\nonumber\\
&+& (e^{\frac{i}{2}(q_1-q_2)\cdot y}-e^{-\frac{i}{2}(q_1+q_2)\cdot y})\nonumber\\ &&\times\int\frac{d^4p}{(2\pi)^4}e^{-ip\cdot y}
{\rm tr}\gamma_{\lambda}\gamma_5 S_{11}(p)\gamma_\rho S_{11}(p+q_2),
\label{div111}
\end{eqnarray}
\begin{eqnarray}
&&(q_1+q_2)_\mu[K_{\mu\rho\lambda}^{(1)112}(q_1,q_2)+K_{\mu\rho\lambda}^{(2)112}(q_1,q_2)]\nonumber\\
&=& (e^{\frac{i}{2}(q_1-q_2)\cdot y}-e^{-\frac{i}{2}(q_1+q_2)\cdot y})\nonumber\\ &&\times\int\frac{d^4p}{(2\pi)^4}e^{-ip\cdot y}
{\rm tr}\gamma_\rho S_{12}(p+q_2)\gamma_{\lambda}\gamma_5 S_{21}(p),
\label{div112}
\end{eqnarray}
\begin{eqnarray}
&&(q_1+q_2)_\mu[K_{\mu\rho\lambda}^{(1)121}(q_1,q_2)+K_{\mu\rho\lambda}^{(2)121}(q_1,q_2)]\nonumber\\
&=&(e^{-\frac{i}{2}(q_1+q_2)\cdot y}-e^{-\frac{i}{2}(q_1-q_2)\cdot y})\nonumber\\ &&\times\int\frac{d^4p}{(2\pi)^4}e^{-ip\cdot y}
{\rm tr}\gamma_\rho S_{21}(p)\gamma_{\lambda}\gamma_5 S_{12}(p+q_1)
\label{div121}
\end{eqnarray}
and
\begin{equation}
(q_1+q_2)_\mu[K_{\mu\rho\lambda}^{(1)122}(q_1,q_2)+K_{\mu\rho\lambda}^{(2)122}(q_1,q_2)]=0
\label{ward122}
\end{equation}
for $a=1$ and similar equations for $a=2$.
We notice that only the momentum integrals that diverges linearly in $y$ as $y\to 0$ contribute to the limit (\ref{limit}). Therefore, the
terms in $S_{ab}(p)$ proportional to the distribution functions can be ignored. Furthermore, the combination of $S_{12}(p+k_1)$ and $S_{21}(p+k_2)$
contributes a product of delta functions, $\delta(p_0+k_{10}+E_{{\bf p}+{\bf k_1}})\delta(p_0+k_{20}-E_{{\bf p}+{\bf k}_2})$ which gives rise to
$\delta(k_{10}-k_{20}+E_{{\bf p}+{\bf k_1}}+E_{{\bf p}+{\bf k}_2})$ which imposes an upper limit of the {\bf p}-integration for a fixed external
momenta $k_1$ and $k_2$ and renders the integral UV finite. Consequently, the only CTP components that contribute to the limit $y\to 0$ of (\ref{dLambda})
correspond to $a=b=c=1$ and $a=b=c=2$ with the vacuum propagators, i.e. $T=\mu=0$. We obtain that
\begin{eqnarray}
(q_1+q_2)_\mu\Lambda_{\mu\rho\lambda}^{111}(q_1,q_2)&=&4(-y_\rho q_{1\mu}\epsilon_{\alpha\mu\beta\lambda}q_{2\beta}-q_2\cdot y\epsilon_{\alpha\rho\beta\lambda}\nonumber\\ &&
+q_1\cdot y\epsilon_{\alpha\rho\beta\lambda}q_{2\beta})u_\alpha(y)
\label{dLambda111}
\end{eqnarray}
and
\begin{eqnarray}
(q_1+q_2)_\mu\Lambda_{\mu\rho\lambda}^{222}(q_1,q_2)&=&4(-y_\rho q_{1\mu}\epsilon_{\alpha\mu\beta\lambda}q_{2\beta}-q_2\cdot y\epsilon_{\alpha\rho\beta\lambda}\nonumber\\ &&
+q_1\cdot y\epsilon_{\alpha\rho\beta\lambda}q_{2\beta})u_\alpha^*(-y)
\label{dLambda222}
\end{eqnarray}
where
\begin{equation}
u_\alpha(y)=\int\frac{d^4p}{(2\pi)^4}\frac{p_\alpha e^{-ip\cdot y}}{(p^2-i0^+)^2}=-\frac{y_\alpha}{8\pi^2 y^2}.
\label{UVintegral}
\end{equation}
Substituting (\ref{UVintegral}) into (\ref{dLambda111}) and (\ref{dLambda222}), we end up with
\begin{eqnarray}
&&(q_1+q_2)_\mu\Lambda_{\mu\rho\lambda}^{111}(q_1,q_2)=(q_1+q_2)_\mu\Lambda_{\mu\rho\lambda}^{222}(q_1,q_2)\nonumber\\
&=&-\frac{1}{2\pi^2}\left(\epsilon_{\mu\beta\lambda\rho}
+\frac{y_\lambda y_\alpha}{y^2}\epsilon_{\rho\alpha\mu\beta}\right)q_{1\mu}q_{2\beta}.
\label{nonconservation}
\end{eqnarray}
where the Schouten identity
\begin{equation}
y_\mu\epsilon_{\rho\lambda\alpha\beta}+y_\beta\epsilon_{\mu\rho\lambda\alpha}+y_\alpha\epsilon_{\beta\mu\rho\lambda}
+y_\lambda\epsilon_{\alpha\beta\mu\rho}+y_\rho\epsilon_{\lambda\alpha\beta\mu}=0
\end{equation}
has been employed. The coordinate representation of (\ref{nonconservation}) reads
\begin{eqnarray}
&&\frac{\partial}{\partial x_\mu}J_\mu(x,y)\nonumber\\=&&\frac{i}{8\pi^2}\left[\epsilon_{\mu\rho\beta\lambda} F_{\mu\rho}(x)F_{5\beta\lambda}(x)
+2\epsilon_{\mu\rho\alpha\beta}\frac{y_\lambda y_\alpha}{y^2}F_{\mu\rho}(x)\frac{\partial}{\partial x_\beta}A_{5\lambda}(x)\right]\nonumber\\
\label{equations}
\end{eqnarray}
where $F_{\mu\nu}=\frac{\partial A_\nu}{\partial x_\mu}-\frac{\partial A_\mu}{\partial x_\nu}$,
$F_{5\mu\nu}=\frac{\partial A_{5\nu}}{\partial x_\mu}-\frac{\partial A_{5\mu}}{\partial x_\nu}$
and the CTP indices have been suppressed. Because of the second term on RHS, the limit $y\to 0$, does not exist in rigorous sense. If we {\it define}
the limit by averaging the direction of $y$ (after continuation to Euclidean space, i.e. $y_0\to -iy_0$), we find \footnote{The anomalous divergence of the
vector current was noted earlier in the context of Kubo formulae \cite{Landsteiner1,Landsteiner2}. The problem here appears more severe since the limit $y\to 0$
is not well-defined rigorously, the direction  average  of $y$ is used . In case of the point-spliting regularization of the chiral anomaly, however, the limit $y\to 0$ is {\it independent} of its direction, as can be shown explicitly
by the Schouten identity.}
\begin{equation}
\frac{\partial}{\partial x_\mu}J_\mu(x)\equiv\frac{\partial}{\partial x_\mu}J_\mu(x,0)
=\frac{3i}{32\pi^2}\epsilon_{\mu\rho\beta\lambda} F_{\mu\rho}(x)F_{5\beta\lambda}(x).
\label{currentdiv}
\end{equation}
It is interesting to note that if the axial potential is a pure gradient, $A_{5\mu}=\frac{\partial\theta}{\partial x_\mu}$, (\ref{equations}) becomes
\begin{eqnarray}
&&\frac{\partial}{\partial x_\mu}J_\mu(x,y)=\frac{i}{4\pi^2}\epsilon_{\mu\rho\alpha\beta}\frac{y_\lambda y_\alpha}{y^2}F_{\mu\rho}(x)\frac{\partial^2\theta}
{\partial x_\beta\partial x_\lambda}
\label{equations}
\end{eqnarray}
and the limit (\ref{currentdiv}) following the hand-waving definition vanishes.

$\ \ \ \ \ \ \ \ \ \ \ \ \ \ \ \ \ \ \ \ \ \ \ \ \ \ \ \ \ \ \ \ \ \ \ \ \ \ \ \ \ \ \ \ \ \ \ \ \ \ \ \ \ \ \ \ \ \ \ \ \ \ \ \ \ \ \ \ \ \ \ \ \ \ $
\noindent{\it 2. An inconsistency}

The electric current, being a functional derivative of the quantum effective action, should satisfy the consistency condition:
\begin{equation}
\frac{\delta J_\mu(x)}{\delta A_\nu(x')}=\frac{\delta J_\nu(x')}{\delta A_\mu(x)}
\end{equation}
which is generalized to
\begin{equation}
\frac{\delta J_\mu^a(x)}{\delta A_\nu^b(x')}=\frac{\delta J_\nu^b(x')}{\delta A_\mu^a(x)}
\end{equation}
in CTP formulation because of the doubling of degrees of freedom. The consistency condition dictates the symmetry property of the current-current
correlator as well as the relationship between the retarded and advanced Green's function of linear response. To the linear order in the external
gauge potential and axial vector potential, the consistency condition requires
\begin{equation}
\lim_{y\to 0}\left[\Lambda_{\mu\rho\lambda}^{abc}(q',q-q')-\Lambda_{\mu\rho\lambda}^{bac}(-q,q-q')\right]=0.
\end{equation}
This is, however, not the case because of the UV divergence.
It follows from (\ref{kernel1}) that
\begin{eqnarray}
&&\Lambda_{\mu\rho\lambda}^{abc}(q',q-q')-\Lambda_{\rho\mu\lambda}^{bac}(-q,q-q')\nonumber\\
&=&-\delta_{ab}[y_\rho K_{\mu\lambda}^{ac}(q-q')-y_\mu K_{\rho\lambda}^{bc}(q-q')]\nonumber\\
&=&-i[K_{\mu\rho\lambda}^{(1)abc}(q',q-q')-K_{\rho\mu\lambda}^{(2)bac}(-q,q-q')\nonumber\\&&+K_{\mu\rho\lambda}^{(2)abc}(q',q-q')-K_{\rho\mu\lambda}^{(1)bac}(-q,q-q')]
\label{diff}
\end{eqnarray}
Upon shifting the integration momentum of $K^{(2)}$, we find that
\begin{eqnarray}
&&K_{\mu\rho\lambda}^{(1)abc}(q',q-q')-K_{\rho\mu\lambda}^{(2)bac}(-q,q-q')\nonumber\\
&=&\left(e^{-\frac{i}{2}q\cdot y}-e^{-\frac{i}{2}q'\cdot y}\right)
\int\frac{d^4p}{(2\pi)^4}e^{-ip\cdot y}\nonumber\\&&\times{\rm tr}\gamma_\mu^aS_{ac}(p+q)\gamma_{\lambda5}^cS_{cb}(p+q')\gamma_\rho^bS_{ba}(p)
\label{incons_1}
\end{eqnarray}
and
\begin{eqnarray}
&&K_{\mu\rho\lambda}^{(2)abc}(q',q-q')-K_{\rho\mu\lambda}^{(1)bac}(-q,q-q')\nonumber\\
&=&\left(e^{\frac{i}{2}q\cdot y}-e^{\frac{i}{2}q'\cdot y}\right)
\int\frac{d^4p}{(2\pi)^4}e^{-ip\cdot y}\nonumber\\&&\times{\rm tr}\gamma_\mu^aS_{ab}(p)\gamma_\rho^bS_{bc}(p-q')\gamma_{\lambda5}^cS_{ca}(p-q).
\label{incons_2}
\end{eqnarray}
Following the arguments after (\ref{ward122}), the components with $a=b=c=1$ and $a=b=c=2$ contribute to a nonzero limit of (\ref{diff}) as $y\to 0$ and the consistency
condition is thereby violated. We obtain that
\begin{eqnarray}
&&\Lambda_{\mu\rho\lambda}^{111}(q',q-q')-\Lambda_{\rho\mu\lambda}^{111}(-q,q-q')\nonumber\\
&=&\Lambda_{\mu\rho\lambda}^{222}(q',q-q')-\Lambda_{\rho\mu\lambda}^{222}(-q,q-q')\nonumber\\
&=&4(y_\alpha\epsilon_{\mu\beta\lambda\rho}+y_\lambda\epsilon_{\rho\alpha\mu\beta})u_\alpha(y)(q-q')_\beta\nonumber\\
&=&-\frac{1}{2\pi^2}\left(\epsilon_{\mu\beta\lambda\rho}+\frac{y_\lambda y_\alpha}{y^2}\epsilon_{\rho\alpha\mu\beta}\right)(q-q')_\beta,
\end{eqnarray}
which implies that
\begin{eqnarray}
&&\frac{\delta J_\mu(x,y)}{\delta A_\nu(x')}-\frac{\delta J_\nu(x,y)}{\delta A_\mu(x)}\nonumber\\
=&&\frac{i}{2\pi^2}\left(\epsilon_{\mu\rho\beta\lambda}+\epsilon_{\mu\rho\alpha\beta}\frac{y_\lambda y_\alpha}{y^2}\right)
\frac{\partial A_{5\lambda}}{\partial x_\beta}\delta^4(x-x^\prime)
\label{incons}
\end{eqnarray}
in coordinate space. Like the case with the current divergence (\ref{equations}), the limit $y\to 0$ does not exists rigorously.
For the limit {\it defined} in (\ref{currentdiv}), we find that
\begin{equation}
\frac{\delta J_\mu(x)}{\delta A_\nu(x')}-\frac{\delta J_\nu(x')}{\delta A_\mu(x)}
=\frac{3i}{16\pi^2}\epsilon_{\mu\rho\beta\lambda}F_{5\beta\lambda}(x)\delta^4(x-x^\prime).
\label{incons1}
\end{equation}
Similar to the current divergence, the first term inside the parentheses on RHS of (\ref{incons}) does not contribute if $A_{5\mu}$ is a pure gradient and the
RHS of (\ref{incons1}) vanishes then.

This inconsistency is related to the nonconservation of the electric current. As the Wigner function is explicitly gauge invariant. The current extracted from it
would be conserved if the current were a functional derivative.

Technically, the process of the limit $y\to 0$ applied to the axial current extracted
from the Wigner function
\begin{eqnarray}
J_{5\mu(x)}&=&i\int\frac{d^4p}{(2\pi)^4}{\rm tr}W(x,p)\gamma_\mu\gamma_5\nonumber\\&=&i\lim_{y\to 0}U(x_+,x_-)<\bar\psi(x_+)\gamma_\mu\gamma_5\psi(x_-)>
\end{eqnarray}
is equivalent to the point-splitting regularization scheme of the axial vector vertex of the triangle diagram in textbooks and this explains why the
the standard axial anomaly is generated from the Wigner function (\ref{wigner}) in \cite{Gao}. However, the same limiting procedure for the electric current (\ref{limit})
amounts to a point-splitting regularization of one of the vector vertex of the same triangle diagram and thereby violates the Bose symmetry. For a non-chiral theory or a
chiral theory with a constant axial vector potential, this is not a problem and the electric current remains conserved and consistent as can be shown explicitly.
In more general situations, a robust regularization scheme has to be introduced to the underlying field theory before defining the Wigner function and the Pauli-Villars
regularization is such a candidate and will be discussed in the next section.

From the regularization perspectives, the dependence of the limit $y\to 0$ on the direction of $y$ may be removed
by inserting a gauge link associated to the axial vector potential, i.e.
\begin{equation}
U_A(x_+,x_-)=e^{i\int_{x_-}^{x_+}d\xi_\mu\gamma_5A_{5\mu}(\xi)}
\end{equation}
into the point-splitted current (\ref{regcurrent}), between $\gamma_\mu$ and $\psi(x_-)$. This amounts to add a term
\begin{equation}
\Delta\Lambda_{\mu\rho\lambda}^{abc}=
-y_\lambda\int\frac{d^4p}{(2\pi)^4}{\rm tr}\gamma_{\mu5}^aS_{ab}(p+q_1)\gamma_\rho^bS_{bc}(p)e^{-i\left(p+\frac{q_1}{2}\right)\cdot y}
\end{equation}
to RHS of (\ref{kernel1}), which in turn cancels the second term on RHS of (\ref{equations}) and the second term on RHS of
(\ref{incons}). Consequently
\begin{equation}
\frac{\partial}{\partial x_\mu}J_\mu(x,y)=\frac{i}{8\pi^2}\epsilon_{\mu\rho\beta\lambda} F_{\mu\rho}(x)F_{5\beta\lambda}(x)
\end{equation}
and
\begin{equation}
\frac{\delta J_\mu(x,y)}{\delta A_\nu(x')}-\frac{\delta J_\nu(x,y)}{\delta A_\mu(x)}
=\frac{i}{4\pi^2}\epsilon_{\mu\rho\beta\lambda}F_{5\beta\lambda}\delta^4(x-x^\prime)
\end{equation}

\section{Chiral magnetic current with regularized Wigner function}

With the Pauli-Villars regularization, a set of Wigner functions pertaining to the regulators, $W(x,p|M_s)$, has to be introduced before the UV unambiguous
electric current $J_\mu(x)$ can be extracted. We have
\begin{eqnarray}
J_\mu(x)&=&ie\int\frac{d^4p}{(2\pi)^4}{\rm tr}[W(x,p)+\sum_sC_sW(x,p|M_s)]\nonumber\\
&=&ie\lim_{y\to 0}U(x_+,x_-)[<\bar\psi_s(x_+)\gamma_\mu\psi_s(x_-)>\nonumber\\&&+\sum_sC_s<\bar\psi_s(x_+)\gamma_\mu\psi_s(x_-)>]
\label{regwigner}
\end{eqnarray}
where the regulator mass $M_s\to\infty$ and $\sum_sC_s=1$. A regulator field, $\psi_s(x)$ is a spinors that generates states with {\it negative norms} in Hilbert space.
The regulators render the limit $y\to 0$ of the quantitiy inside the bracket on RHS of (\ref{regwigner}) finite. Therefore the gauge link $U(x_+,x_-)$ can be set to one in
what follows and we end with
\begin{equation}
J_\mu(x)=-ie{\rm tr}\gamma_\mu<\psi(x)\bar\psi(x)>-ie\sum_sC_s{\rm tr}\gamma_\mu<\psi_s(x)\bar\psi_s(x)>
\end{equation}
In another word, replacing the massless fermion propagators in (\ref{kernel1}) with massive ones will not alter its UV behavior. Combining (\ref{kernel1}) with
the counterpart from the PV regulators, the UV divegence cancel and the limit $y\to 0$ becomes trivial.
As is shown in the last section, $J_\mu(x)$ can be written in terms of the CTP propagator ${\cal S}(x,y)$ in the presence of external $A_\mu$ and $A_{5\mu}$,
\begin{equation}
J_\mu(x)=-ie\frac{1}{2}\left[{\rm Tr}\gamma_\mu{\cal S}_0(x,x)-\sum_sC_s{\rm Tr}\gamma_\mu{\cal S}_s(x,x)\right]
\end{equation}
with ${\rm Tr}$ acting on both CTP an spinor indices. The rest of the caculation is purely field theoretic but without assuming particular form of the one-particle
distribution function.
In place of (\ref{JLambda1}), (\ref{kernel1}) and (\ref{kernel2}), we have
\begin{eqnarray}
J_\mu(x)=&&e^2\int\frac{d^4q_1}{(2\pi)^4}\int\frac{d^4q_2}{(2\pi)^4}e^{i(q_1+q_2)\cdot x}\nonumber\\&&\times\Lambda_{\mu\rho\lambda}^{\rm Reg.}(q_1,q_2)
A_\rho(q_1)A_{5\lambda}(q_2)
\end{eqnarray}
where the regularized kernel
\begin{eqnarray}
&&\Lambda_{\mu\rho\lambda}^{\rm Reg.}(q_1,q_2)\nonumber\\&=&\int\frac{d^4p}{(2\pi)^4}\left[I_{\mu\rho\lambda}(p,q_1,q_2|0)-\sum_{s>0}C_sI_{\mu\rho\lambda}(p,q_1,q_2|M_s)\right]\nonumber\\
\label{integration}
\end{eqnarray}
where
\begin{eqnarray}
&&I_{\mu\rho\lambda}(p,q_1,q_2|m)\nonumber\\
&\equiv&-\frac{i}{2}{\rm Tr}[\gamma_\mu S(p+q_1+q_2|m)\Gamma_\lambda\gamma_5S(p+q_1|m)\Gamma_\rho S(p|m)\nonumber\\
&&+\gamma_\mu S(p+q_1+q_2|m)\Gamma_\rho S(p+q_2|m)\Gamma_\lambda\gamma_5S(p|m)]\nonumber\\
\label{integrandI}
\end{eqnarray}
with $S(p|m)$ the CTP propagator of a massive spinor and $\Gamma_\mu={\rm diag.}(\gamma_\mu,-\gamma_\mu)$. Because of the triviality of the limit $y\to 0$,
the right hand sides of (\ref{div111}), (\ref{div112}), (\ref{div121}), (\ref{incons_1}) and (\ref{incons_2}) all vanish with PV regularized kernels
$K^{(1)}$ and $K^{(2)}$. So do the right hand sides of (\ref{divLambda}) and (\ref{diff}) then, and we end up with a conserved and consistent electric
current with an arbitrary external axial vector potential $A_{5\mu}(x)$. In another word, the contribution of the Wigner functions associated to the PV regulators
play the role of the Bardeen like terms that removes the current divergence and incnsistency.

The case of a constant axial chemical potential with the PV regularized Wigner function is, however, rather subtle. It depends on how the limit
$q_2\to 0$ is taken and two different orders of the limit are calculated below.

Consider the order $\lim_{q_{20}\to 0}\lim_{\vec q_2\to 0}$ first. Upon setting $\vec q_2=0$, we have $q\!\!\!/_2=q_{20}\gamma_4$
\begin{eqnarray}
&&q_{20}I_{\mu\rho4}(p,q_1,q_2|m)\nonumber\\ &=& -i{\rm Tr}[\gamma_\mu S(p+q_1+q_2|m) Q\!\!\!/_2\gamma_5S(p+q_1|m)\Gamma_\rho S(p|m)\nonumber\\
&&+\gamma_\mu S(p+q_1+q_2|m)\Gamma_\rho S(p+q_2|m)Q\!\!\!/_2\gamma_5S(p|m)]\nonumber\\
&=& {\rm Tr}\gamma_\mu[\gamma_5S(p+q_1|m)\Gamma_\rho S(p|m)\nonumber\\&&-\gamma_5S(p+q_1+q_2|m)\Gamma_\rho S(p+q_2|m)]+D_{\mu\rho}(p,q_1,q_2|m)\nonumber\\
\label{identity}
\end{eqnarray}
where
\begin{eqnarray}
&&D_{\mu\rho}(p,q_1,q_2|m)\nonumber\\&\equiv&-im{\rm Tr}\gamma_\mu[S(p+q_1+q_2|m)\Gamma_5S(p+q_1|m)\Gamma_\rho S(p|m)\nonumber\\
&&+S(p+q_1+q_2|m)\Gamma_\rho S(p+q_1|m)\Gamma_5S(p|m)]
\end{eqnarray}
with $\Gamma={\rm diag.}(\gamma_5,-\gamma_5)$
and the identity
\begin{eqnarray}
S(p+q|m)Q\!\!\!/\gamma_5S(p)&=&i\left[\gamma_5S(p|m)+S(p+q|m)\gamma_5\right]\nonumber\\&&+2mS(p+q|m)\Gamma_5S(p|m)
\end{eqnarray}
is employed. The two terms inside the first brackets of (\ref{identity}) differs by a shift of the integration momentum $p$ and their contribution to the integration
cancel each other once regularized. Because $D_{\mu\rho}(p,q_1,q_2|0)=0$, we are left with only the regulator contribution to the kernel, i.e.
\begin{equation}
q_{20}\Lambda_{\mu\rho4}^{\rm Reg.}(q_1,q_2)=-\int\frac{d^4p}{(2\pi)^4}\sum_sC_sD_{\mu\rho}(p,q_1,q_2|M_s)
\end{equation}
In the limit, $M_s\to\infty$, the distribution functions $f(E)$ pertaining to the CTP propagator vanishes, and the terms of the structure $...S_{12}(p+k_1)...S_{21}(p+k_2)...$ or
$...S_{21}(p+k_1)...S_{12}(p+k_2)...$ are proportional to $\delta(k_{10}-k_{20}+E_{\vec p+\vec k_1}+E_{\vec p+\vec k_2})=0$. Furthermore, the terms with the same CTP indices, 1 or 2
contribute equally, where the negative sign pertaining to the anti-time ordering is compensated by the negative sign of the Wick rotation in this case. Therefore
$D_{\mu\rho}(p,q_1,q_2|m)$ in the integrand of (\ref{integration}) can be replaced by the expression
\begin{eqnarray}
&&-2m{\rm tr}\gamma_\mu[\frac{1}{p\!\!\!/+q\!\!\!/_1+q\!\!\!/_2-m}\gamma_5\frac{1}{p\!\!\!/+q\!\!\!/_1-m}\gamma_\rho\frac{1}{p\!\!\!/+q\!\!\!/_1-m}\nonumber\\
&&+\frac{1}{p\!\!\!/+q\!\!\!/_1+q\!\!\!/_2-m}\gamma_5\frac{1}{p\!\!\!/+q\!\!\!/_1-m}\gamma_\rho\frac{1}{p\!\!\!/+q\!\!\!/_1-m}]\nonumber\\
&=&4m^2q_{20}\frac{{\rm tr}\gamma_5\gamma_\mu q\!\!\!/_1\gamma_4}{p^2+m^2}+O(q_{20}^2).
\end{eqnarray}
Consequently, we obtain for the CME that
\begin{eqnarray}
&&\lim_{q_{20}\to 0}\lim_{\vec q_2\to 0}\Lambda_{ij4}^{\rm Reg.}(q_1,q_2)\nonumber\\=&&-16i\epsilon_{ikj}q_{1k}\int\frac{d^4p}{(2\pi)^4}\sum_{s>0}\frac{C_sm_s^2}{(p^2+M_s^2)^3}
\nonumber\\=&&-\frac{1}{2\pi^2}\epsilon_{ikj}q_{1k}
\label{order1}
\end{eqnarray}
which generates the CME current.

To evaluate the limit with the opposite order, i.e., $\lim_{\vec q_2\to 0}\lim_{q_{20}\to 0}$, we set $q_2=(0,\vec q_2)$. As was pointed out in \cite{Hou}, the difference
from limit order (\ref{order1}) stems from coalescence of the poles of the two propagators adjacent to $\Gamma_\lambda\gamma_5$ in (\ref{integrandI}). In terms of the
retarded (advanced) propagator $S_R$ ($S_A$) and the correlator $S_c$, defined via
\begin{equation}
S(p)=Q^{-1}\left(\begin{array}{cc} 0 & S_R(p)\\
                               S_A(p) & S_c(p)\\ \end{array}\right)Q
\end{equation}
with
\begin{equation}
Q=\frac{1}{\sqrt{2}}\left(\begin{array}{cc} 1 & -1\\
                                      1 & 1\\ \end{array}\right).
\end{equation}
we find that
\begin{widetext}
\begin{align}
S(p+q)\Gamma_4\gamma_5S(p)
=Q^{-1}\left(\begin{array}{cc} 0 & S_A(p+q)\gamma_4\gamma_5S_A(p)\\
                                      S_R(p+q)\gamma_4\gamma_5S_R(p) & S_c(p+q)\gamma_4\gamma_5S_A(p)+S_c(p+q)\gamma_4\gamma_5S_R(p)\\ \end{array}\right)Q
\end{align}
\end{widetext}

The matrix elements with exclusively $S_R$ or $S_A$ are not sensitive to the orders of limit $q\to 0$ since integration contour on the $p_0$ plane can always
be deformed away from the poles and we need only to focus our attention to the part

\begin{eqnarray}
&&Q^{-1}\left(\begin{array}{cc} 0 & 0\\
                              0 & S_c(p+q)\gamma_4\gamma_5S_A(p)+S_c(p+q)\gamma_4\gamma_5S_R(p)\\ \end{array}\right)Q\nonumber\\
\equiv&& {\cal P}(p,q)\left(\begin{array}{cc} 1 & 1\\
                                      1 & 1\\ \end{array}\right)
\end{eqnarray}
where
\begin{eqnarray}
{\cal P}(p,q)&=&\frac{1}{2}\left[S_c(p+q)\gamma_4\gamma_5S_A(p)+S_R(p+q)\gamma_4\gamma_5S_c(p)\right]\nonumber\\
&=&-i\pi D(p,q)(p\!\!\!/+q\!\!\!/)\gamma_4\gamma_5p\!\!\!/
\end{eqnarray}
with
\begin{equation}
D(p,q)\equiv\frac{[1-2f(|\vec p+\vec q|)]\delta[(p+q)^2]}{p^2}+\frac{[1-2f(|\vec p|)]\delta(p^2)}{(p+q)^2}
\end{equation}
On writing the kernel in the form
\begin{eqnarray}
&&\Lambda_{ij4}^{\rm Reg.}(q_1,q_2)\nonumber\\&=&\int\frac{d^4p}{((2\pi)^4}[D(p+q_1,q_2)U_{ij}(p_0+q_{10},\vec p+\vec q_1|q_1,q_2)
\nonumber\\&+&D(p,q_2)U_{ij}^\prime(p_0,\vec p|q_1,q_2)]\nonumber\\&+&\hbox{the terms not sensitive to the orders of limits}
\end{eqnarray}
with
\begin{eqnarray}
&&U_{ij}(p_0+q_{10},\vec p+\vec q|q_1,q_2)\nonumber\\&=&-\pi{\rm tr}\gamma_i(p\!\!\!/+q\!\!\!/_1+q\!\!\!/_2)\gamma_4\gamma_5(p\!\!\!/+q\!\!\!/_1)\gamma_jS_A(p)\\
&&U_{ij}^\prime(p_0,\vec p|q_1,q_2)\nonumber\\&=&-i\pi{\rm tr}\gamma_iS_R(p+q_1+q_2)\gamma_j(p\!\!\!/+q\!\!\!/_1)\gamma_4\gamma_5p\!\!\!/
\end{eqnarray}
the difference between the two orders of the limit $q_2\to 0$ is given by \cite{Hou}
\begin{eqnarray}
& &\lim_{\vec q_2\to 0}\lim_{q_{20}\to 0}\Lambda_{ij4}^{\rm Reg.}(q_1,q_2)-\lim_{q_{20}\to 0}\lim_{\vec q_2\to 0}\Lambda_{ij4}^{\rm Reg.}(q_1,q_2)\nonumber\\
&=&\frac{1}{2}\int\frac{d^3\vec p}{(2\pi)^3}[\frac{df(|\vec p'|)}{d|\vec p'|}\frac{U_{ij}(|\vec p'|,\vec p'|q_1,0)+U_{ij}(-|\vec p'|,\vec p'|q_1,0)}{|\vec p'|^2}\nonumber\\
&+&\frac{df(|\vec p|)}{d|\vec p|}\frac{U_{ij}^\prime(|\vec p|,\vec p|q_1,0)+U_{ij}(-|\vec p|,\vec p|q_1,0)}{|\vec p|^2}]
\end{eqnarray}
with $p'\equiv p+q_1$.
It is straightforward to show that
\begin{eqnarray}
\frac{U_{ij}(|\vec p'|,\vec p'|q_1,0)}{|\vec p'|^2}&=&\frac{U_{ij}(-|\vec p'|,\vec p'|q_1,0)}{|\vec p'|^2}\nonumber\\&=&-8P\frac{1}{\vec p^2-\vec p^{\prime2}}\epsilon_{ikj}q_{1k}\nonumber\\
\frac{U_{ij}^\prime(|\vec p|,\vec p|q_1,0)}{|\vec p|^2}&=&\frac{U_{ij}^\prime(-|\vec p|,\vec p|q_1,0)}{|\vec p|^2}\nonumber\\&=&8P\frac{1}{\vec p^2-\vec p^{\prime2}}\epsilon_{ikj}q_{1k}
\end{eqnarray}
with $P(...)$ standing for the principal value upon integration. Consequently,
\begin{eqnarray}
&&\lim_{\vec q_2\to 0}\lim_{q_{20}\to 0}\Lambda_{ij4}^{\rm Reg.}(q_1,q_2)\nonumber\\ &=& \lim_{q_{20}\to 0}\lim_{\vec q_2\to 0}\Lambda_{ij4}^{\rm Reg.}(q_1,q_2)
+4P\int\frac{d^3\vec p}{(2\pi)^3}\frac{\frac{df(|\vec p'|)}{d|\vec p'|}-\frac{df(|\vec p|)}{d|\vec p|}}{\vec p^{\prime2}-\vec p^2}\nonumber\\
&=& -\frac{1}{2\pi^2}\epsilon_{ikj}q_{1k}+8\epsilon_{ikj}q_{1k}P\int\frac{d^3\vec p}{(2\pi)^3}\frac{df(|\vec p|)}{d|\vec p|}\frac{1}{\vec p^2-(\vec p+\vec q_1)^2}\nonumber\\
&=& \frac{2f(0)-1}{2\pi^2}\epsilon_{ikj}q_{1k}+O(q_1^2)
\end{eqnarray}
The CME current is canceled at thermal equilibrium where $f(|\vec p|)=\frac{1}{e^{\beta|\vec p|}+1}$, consistent with the statement in the literature \cite{Hou}.

Note that throughout the calculation, we never relied on the explicit form of the distribution function $f(|\vec p|)$ other than the limiting behavior
$f(\infty)=0$. Therefore, the result
obtained in this section is more general than that follows from the Matsubara formulation.

\section{Concluding remarks}

In summary, we point out some problems of the Wigner function formalism in its present form because of the UV divergence of the underlying quantum field
theory. We find that in the presence of a non-constant electromagnetic gauge potential and a non-constant axial vector potential, the electric current
extracted from the present form of the Wigner function defined by (\ref{wigner}) is neither consistent (a functional derivative of some effective action with respect to the
gauge potential) nor conserved.The issue is closely related to UV ambiguities associated to the axial anomaly. Therefore, the present form of the Wigner function
formulation needs to be refined to be applicable to a general chiral plasma and a robust regularization scheme needs to be introduced.

Then we explored the Pauli-Villars regularization scheme by introducing the Wigner functions of the regulators, which remove the ambiguity of the limit $y\to 0$ in
(\ref{limit}) and restores the current conservation and consistency with an arbitrary vector and axial vector potentials. In case of an inhomogeneous and time dependent
axial chemical potential, the constant limit depends its order. The full chiral magnetic current is recovered if the spatial inhomogeneity is removed first but a different chiral magnetic current emerges if
the time dependence is turned off first, which vanishes at thermal equilibrium. This result may have phenomenological implications.

Including the QCD interactions, the Wigner function formalism will get more complicated with UV renormalization of QCD vertices. The Wigner functions of the PV regulators,
introduced in this work, may be intact since they are tied to the axial anomaly and will not be renormalized.

\begin{acknowledgments}

We are indebted to H. Liu for her early participation of this project. We thanks K. Fukushima, S. Lin, L. McLerran and Q. Wang for helpful discussions. The interesting comments of K.
Landsteiner, M. Chernodub and O. Ruchayskiy are warmly acknowledged. This research is partly supported by the Ministry of Science and Technology of China (MSTC) under the "973" Project No. 2015CB856904(4).   D-f. Hou and H-c. Ren are partly supported by NSFC under Grant Nos. 11375070, 11521064£¬11135011. Y. Wu is supported by the Fundamental Research Funds for the Central China Normal Universities, and the QLPL under grant No. 201507.
\end{acknowledgments}

\newpage 
\end{document}